\documentclass[12pt]{iopart}

\usepackage{graphicx}  
\usepackage{epsfig}
\begin{document}

\title[Out of equilibrium dynamics]{The out of equilibrium dynamics of 
the Sherrington-Kirkpatrick model}

\author{Leticia F. Cugliandolo$^1$ and Jorge Kurchan$^2$}

\address{
$^1$
Universit\'e Pierre et Marie Curie - Paris VI,\\
Laboratoire de Physique Th\'eorique et Hautes Energies, \\
4. Place Jussieu, 75252 Paris Cedex 05 France\\
$^2$ PMMH, ESPCI rue Vauquelin, 75251 Paris Cedex 05 France}
\ead{leticia@lpthe.jussieu.fr; jorge@pmmh.espci.fr}
\begin{abstract}
  The analytic solution to the  dynamics of the 
Sherrington-Kirkpatrick model was developed in the
nineties. It involves
directly measurable out of equilibrium quantities,
and thus addresses  the  questions relevant to an experimental
system.  We here review the out of equilibrium relaxation of this 
model and how it compares to experimental measurements.
\end{abstract}

\maketitle

\section{Introduction}
\label{sec:introduction}

 Most analytic studies of spin-glasses carried out before the early
90s
 focused on their Gibbs-Boltzmann measure.
 The use of the
replica
 trick, the cavity method and the Thouless-Anderson-Palmer
(TAP)
 approach yielded a
 rather complete description of the
equilibrium states
 of the
 Sherrington-Kirkpatrick model~\cite{SK}
and other disordered systems. The picture
 that emerged is one of
an
 extremely complex free-energy landscape with many minima, the
lowest
 of which are the equilibrium `pure states'.  
 The
geometrical organisation of these
 states, and their relative weights
in the equilibrium measure
 are the main objects in the Parisi
theory~\cite{Mepavi}, 
 at the centre of which is the
 functional
order parameter $P(q)$ giving the probability that two equilibrium
configurations in a randomly chosen sample
 have an overlap $q$.

A much more difficult programme, not quite completed yet, concerns
the
 understanding of the
 organisation of the landscape away from
the equilibrium configurations.
 Questions such as the metastable
state stability, their basins of
 attraction, and the nature of the

 barriers separating them, have proven to be much harder to answer
in an unambiguous way.  
 This more detailed knowledge of the
landscape may seem a necessary
 condition for the understanding of
the experimental, non-equilibrium situation.
 Surprisingly enough, it
turns out that a direct solution of the 
 out of equilibrium dynamics
is in fact quite simpler.

The dynamic approach was pioneered by Sompolinsky and Zippelius in the
early 80's, as a method to avoid the use of replicas for the calculation of
equilibrium quantities \cite{Sozi}. They introduced the
general framework and succeeded in calculating the 
high-temperature quantities. The low-temperature situation turned out to be 
more complicated, and the problem  of finding  
a true dynamic equilibrium solution  remains open to this day.

An alternative approach, developed in the early 90's, is to study the out of
equilibrium dynamics starting from a quench in temperature, just as in the
experimental protocols~\cite{Cuku93,Cuku94}. Although one might have expected that such 
a situation is hopelessly difficult, as it
involves  the landscape  
far from  equilibrium, including  the   basins of attraction  and
barriers associated with metastable states, it turns out that
the actual analytic solution is only slightly -- if at all -- harder than the
equilibrium one using replicas. 

In general, the Gibbs-Boltzmann measure can be explored with a
stochastic process that satisfies detailed balance. In order 
for the system to equilibrate,
   the limit of
 large times taken {\it before} the thermodynamic limit:
\begin{equation}
\lim_{N \rightarrow \infty} \;\;\; \lim_{t \rightarrow \infty}
\; . 
\label{lim0}
\end{equation}
Times are measured after a preparation instant, typically the moment
when an instantaneous quench into the high or low-temperature phase is
performed and temperature is henceforth kept constant.  The order of
limits (\ref{lim0}) guarantees ergodicity since barriers can be
overcome at sufficiently long times for finite $N$.  The equilibrium
thermodynamical values of any operator $O$ are then obtained as the
long-time limit of noise averaged time-dependent
observables,
$\langle \,  O \, \rangle_{eq}=
\lim_{N\rightarrow\infty} \lim_{t\rightarrow\infty}  \langle \, O(t) \, \rangle 
$.

The existence of divergent barriers in spin-glass mean-field models
led Sompolinsky~\cite{So} to postulate that these systems relax in a
set of   hierarchically ordered 
 time scales that eventually diverge with $N$. 
These $N$-dependent timescales entered the  solution proposed 
for the saddle-point
equations of motion via the time-decay of the correlation and
response functions. In this {\it Ansatz},
 although equilibrium was assumed, the 
the fluctuation-dissipation relation between correlation and response
 was violated.
 As several authors pointed out, this  is clearly inconsistent 
~\cite{Mepavi,Hojayo,Frku}:  the problem can be traced 
back to  the fact that  the saddle point dynamic equations are only valid 
when $N \to \infty$ and the times are kept finite.
In any event, both the existence of many timescales and 
the important role of the fluctuation-dissipation relations were found to be
 crucial 
features of the problem that reappeared in later developments.

A different situation, closer to the experimental procedure, is to
consider the relaxation of infinite systems at long but finite times
using initial conditions that are not correlated with the quenched
disorder~\cite{Cuku93}.  The order of limits is then
\begin{equation}
\lim_{t \rightarrow \infty} \;\;\; \lim_{N \rightarrow \infty}
\label{lim1}
\; .
\end{equation}
Divergent barriers in the thermodynamic limit 
imply ergodicity-breaking: the relaxational
dynamics does not explore
the full phase-space in finite times at large $N$. In fact,
there is no equilibration time $t_{eq}$ such that for all subsequent times
the system reaches either the Gibbs-Boltzmann distribution or any
time-independent distribution in a fixed, restricted sector of phase
space. The dynamics is for all times something different from local
equilibrium.
This is the phenomenon of {\em  aging}: the relaxation of the system
depends on its history at all  times.  Though aging
effects lie beyond the scope of thermodynamics, they have
been observed in numerous disordered systems.  As we shall see below,
 the dynamics of
mean-field disordered models (\ref{lim1}) capture aging
phenomena with similarities and differences from what is observed
experimentally.

In what follows we  summarise what is known about the out of
equilibrium dynamics of the Sherrington-Kirkpatrick (SK) model~\cite{SK}. We
describe the analytic solution to the relaxational dynamics in the
limit (\ref{lim1})~\cite{Cuku94,Bacukupa95}
 and we briefly confront its behaviour to the one
observed in experimental systems.

\section{The Sherrington-Kirkpatrick (SK) model}
\label{sec:themodel}

The SK Hamiltonian is $H =-\sum_{i < j}^N J_{ij} s_i s_j$ where
the interaction strength $J_{ij}$ are
independent random variables with a Gaussian distribution with zero mean
and variance
$[J_{ij}^2]=
1/(2 N)
$.
The square brackets stand for the average over the couplings.
The spin variables take values $\pm 1$~\cite{SK}. 

Although the natural dynamics 
for Ising spin systems are of Glauber~\cite{Szamel} or
Montecarlo type, these are not well adapted to implement analytical
calculations. It is then preferable to transform the discrete
variables into continuous ones and to use Langevin dynamics~\cite{Sozi}. The
Hamiltonian of the soft-spin SK model is then
\begin{equation}
H =
-
\sum_{i < j}^N
J_{ij}
s_i s_j
+ a \sum_i (s_i^2-1)^2 +
\frac{1}{N^{r-1}}
\sum_{i_1 < \dots < i_r}^N
h_{i_1  \dots i_r}
s_{i_1} \dots s_{i_{r}}
\; ,
\end{equation}
$-\infty \leq s_i \leq \infty$, $\forall i$.
Letting $a \rightarrow \infty$ one recovers the Ising case, although this is
not essential.
 Additional
source terms ($h_{i_1 \dots i_{r}}$ time-independent)
have been included. If $r=1$ the Zeeman coupling to a local magnetic
field $h_i$ is recovered. 

The dynamics is given by the Langevin equation
\begin{equation}
\Gamma_0^{-1} \, \partial_t \sigma_i(t)
=
- \frac{\delta H}{\delta \sigma_i(t)}
+ f_i(t) +\xi_i(t)
\;
\label{lang}
\end{equation}
$\Gamma_0$ determines the time scale and it is henceforth set to one. 
$\xi_i(t)$ is a Gaussian white noise with zero mean and variance $2k_BT$
and we set $k_B=1$ hereafter.
$f_i(t)$ represent any other perturbing force. For example, non-potential 
forces are important in the analysis of rheological experiments
and are mimicked as
\begin{equation} 
f_i = \epsilon \sum_{j\neq i} J^a_{ij} s_j  
\label{antisim}
\end{equation}
with $J_{ij}^a$ an antisymmetric matrix, $J_{ij}^a=-J_{ji}^a$.  Forces
that oscillate in time can be used  to mimic 
the slow relaxation under shaking of  systems such as
 granular matter.  Such forces maintain the system in a driven
out of equilibrium regime even if the limit (\ref{lim0}) is
considered.  The mean over the thermal noise is hereafter represented
by $\langle \;\dots\;\rangle$.


The dynamics of a Langevin process is usually expressed with a functional
integral for the generating functional by using the so-called
Martin-Siggia-Rose method.  As De Dominicis first pointed out,
 one does not need to use the replica trick to analyse
the relaxation of models with quenched disorder if the initial 
condition is not correlated with the quenched randomness~\cite{deDo}.
 The analysis of the relaxational dynamics
of a disordered model is thus considerably more straightforward
 than that of   the statics, in particular since all observables have a 
very clear physical interpretation and can be easily and directly accessed
with experiments and numerical simulations.

The sample-averaged dynamics for $N \rightarrow \infty$ is entirely
described by the evolution of the two-time correlation and the linear
response functions \cite{Sozi}
\[
C(t,t')
\equiv
\frac{1}{N} \sum_{i=1}^N [\langle s_i(t) s_i(t') ]
\qquad\qquad
R(t,t')
\equiv
 \frac{1}{N}
\sum_{i=1}^N
\left.
\left[
\frac{\delta \langle s_i(t) \rangle}
     {\delta h_i(t')}\right|_{h=0}
\right]
\; .
\]
The square brackets denote disorder average.
Exact dynamic equations for the evolution of these dynamic macroscopic
order parameters, in the {\it strict} large $N$ limit taken at the
outset of the calculation, have been written down by Sompolinsky and
Zippelius~\cite{Sozi}. They are rather cumbersome because, just as in
the static case, the spin variables cannot be explicitly integrated
away.  Several paths can be followed to approximate the effect of the
quartic term introduced by the soft-spin potential.  One possibility
is to use a mode-coupling approximation~\cite{Kech}.  Another
possibility is to focus on the dynamics close to the critical
temperature, use the fact that the transition is expected to be second order, 
and deal with the dynamic counterpart of the `truncated
model' introduced  by Parisi for the equilibrium case. 

 All parameters in the
resulting large-$N$ equations are independent of $N$ and finite,
and have a unique solution.
In the high temperature, $T>T_g=1$, regime the evolution
reaches equilibrium, while below $T_g$ this is no longer the case.  In
the following we focus on the relaxation in the low-temperature phase.

\section{Analytic solution}
\label{sec:analytic}

In this Section we summarise the analytic solution to the SK model~\cite{Cuku94}.

\subsection{Generic properties}

The following properties appear to be quite generic of 
glassy dynamics. 

\vspace{.2cm}

{\em Separation of time-scales.} 

\vspace{.2cm}

After a (long) time $t'$
there is a quick relaxation in a `short' time-delay $t-t'$ and the
self correlation decays to a value $q_{ea}$, followed by a slower
drift away. The parameter $q_{ea}$ is interpreted as the
Edwards-Anderson parameter that represents the size of a `trap' or the
`width of a channel' in phase space. Within these traps the system is
fully ergodic while it becomes more and more difficult to escape a
trap as time passes.  The correlation and response functions can thus be
written in a way that explicitly separates the terms corresponding to
the relaxation within a trap:
\begin{equation}
C(t,t')=C_{st}(t,t')+C_{ag}(t,t')
\; ,
\qquad
R(t,t')=R_{st}(t,t')+R_{ag}(t,t')
\; .
\label{fdt}
\end{equation}
Consistently, $C_{st}(t,t')$ and $R_{st}(t,t')$ are assumed
to satisfy the equilibrium relations, {\it i.e.} time homogeneity and the
fluctuation-dissipation theorem (FDT)
\begin{eqnarray}
\begin{array}{rcl}
C_{st}(t,t') &=& C_{st}(t-t')
\nonumber\\
R_{st}(t,t') &=& R_{st}(t-t')
\end{array}
\;\;\;\;\;\;\;\;\;
R_{st}(t-t')&=& \theta(t-t') \frac{\partial C_{st}(t-t')}{\partial t'} \; ,
\end{eqnarray}
and
\begin{eqnarray}
\begin{array}{rclrcl}
C_{st}(0)&=&1-q_{ea}
\; ,
&
\;\;\;\;\;\;
\lim_{t-t' \rightarrow \infty} C_{st}(t-t') &=& 0
\; ,
\nonumber\\
C_{ag}(t,t) &=& q_{ea}
\; ,
&
\lim_{t \rightarrow \infty} C_{ag}(t,t')&=&0
\; .
\end{array}
\end{eqnarray}

\vspace{.2cm}

{\em Weak ergodicity-breaking.} 

\vspace{.2cm}

The correlation satisfies:
\begin{equation}
\frac{\partial C_{ag}(t,t')}{\partial t} \leq 0
\; , 
\qquad
\frac{\partial C_{ag}(t,t')}{\partial t'} \geq 0
\; .
\label{web}
\end{equation}
This means that the system, after a given time $t'$, starts drifting away
(albeit slowly) until it reaches the maximal
distance (in general much larger than the size of a state)
 at sufficiently long times $t$.

\vspace{.2cm}

{\em Weak long-term memory.} 

\vspace{.2cm}

The integrated linear response satisfies:
\begin{equation}
\lim_{t \rightarrow \infty} \chi (t,t') \equiv 
\lim_{t \rightarrow \infty} \int_{0}^{t'} \; dt'' \;
R(t,t'')
=0 \;\;\;\; \forall \;
\mbox{fixed} \; t'
\; .
\label{wltm}
\end{equation}
$\chi (t,t')$ is the normalised (linear) response at time $t$ to a
constant small magnetic field applied from $t''=0$ up to $t''=t'$,
often called the `thermoremanent magnetisation'.  This hypothesis is
crucial, since the response function represents the memory the
system has of what happened at previous times: the weakness of the
long-term memory implies that the system responds to its past in an
averaged way, the details of what takes place during a finite time
tend to be washed away (the `high-school' effect).
In interesting cases, the system however does not have only short-term memory:
\begin{equation}
\lim_{t \rightarrow \infty}  \int_{0}^{t} \; dt' \;
R_{ag}(t,t') > 0 
\label{wltm1}
\end{equation}
so that fields acting during an appreciable
fraction of the  distant past have a finite effect.

\subsection{High frequency `quasi-equilibrium' dynamics}

The Sompolinsky-Zippelius~\cite{Sozi} results for the equilibrium dynamics 
within a pure state can be reinterpreted in the out of equilibrium 
context to describe the first quick relaxation regime. One finds that the 
self-correlation decays to the plateau at the 
Edwards-Anderson parameter given by 
\begin{equation}
(\tau-q_{ea})+q_{ea}^2=0 \, \qquad \Rightarrow \qquad 
q_{ea} = \frac{1+\sqrt{1-4\tau}}{2} \sim 1-\tau 
\end{equation}
 where $\tau=T_g-T$.
\begin{equation}
C_{st}(t,t') \sim (1-q_{ea}) + A (t-t')^{-a(T)}
\end{equation}
with $a(T)$ a nontrivial temperature-dependent exponent. 

\subsection{Aging regime}

The dynamic equations in the aging regime can be solved by using the 
properties listed above. The slowness of the dynamics allows
one to neglect the effect of the time-derivative and write down coupled 
integral equations for $C_{ag}$ and $R_{ag}$. These equations are 
invariant under reparametrisations of time, $t\to h(t)$,  that transform 
the `fields' $C_{ag}$ and $R_{ag}$ as
\begin{equation}
C_{ag}(t,t') \to C_{ag}(h(t),h(t'))
\; , 
\qquad
R_{ag}(t,t')  \to \dot h(t') R_{ag}(h(t),h(t'))
\label{eq:RpG}
\end{equation} 

This invariance deserves some explanation. The complete equations
of motion have no such symmetry. However, in the large-time
limit, the equations for the `aging' correlation and responses $C_{ag},R_{ag}$
become less and less dependent of the timescale, because time-derivative
terms become less and less relevant. Only in the infinite time limit are 
time-derivatives negligible and 
reparametrisations become a true symmetry.

 (Broken) symmetries are  related to divergent susceptibilities
and large spontaneous fluctuations.
Here, these susceptibilities and  fluctuations will diverge
only in the large-time (or vanishing frequency) limit.

 Let us see this with a concrete example~\cite{Babeku}. 
If we add to the equations of motion
(\ref{lang}) a forcing term of the form (\ref{antisim}), one can show that
aging disappears {\em whatever the value of $\epsilon$}, and correlations
become stationary at large times:
\begin{equation}
C_{ag}(t,t') = {\cal C}\left( \frac{\ln(t-t')}{\epsilon}\right)
\label{ludovic}
\end{equation}
The time for such a stationary regime to be achieved grows with
$\epsilon$.
 Hence, we have that the system has an arbitrarily large
susceptibility 
 with respect to the forces $f_i$ (because the
two-point correlation
 functions depend strongly on them), provided
we wait long enough.
 
 One can also reason in terms of spontaneous
fluctuations, as we shall
 see below, a path that suggests the
introduction of a `sigma model'
 that encapsulates the fluctuations
in the `almost-flat' directions~\cite{fluct-review}.

\subsection{Correlation scales}

The analysis of the aging regime in the SK model motivated the study
of generic properties of time correlation functions and the
development 
of a complete classification of their possible behaviour~\cite{Cuku94}.

Take three ordered times $t_3 \geq t_2 \geq t_1$, and the corresponding
correlations
are $C(t_i,t_j)$
The monotonicity of the decay of the correlations with respect to the 
longer time (keeping the shorter time fixed) and the shorter time (keeping the 
longer time fixed) allows us to derive general properties 
that strongly constrain the possible scaling forms. Indeed, one 
can relate any three
correlation functions via {\it triangle relations}~\cite{Cuku94} 
\begin{eqnarray}
\lim_{\stackrel{ 
t_1 \to\infty}{\stackrel{
C(t_2,t_1)=C_{21}}{C(t_3,t_2)=C_{32}}} 
}
C(t_3,t_1) 
&=& f(C_{32},C_{21})
\; .
\end{eqnarray}
where $f(x,y)$ is a function that determines the form of the 
 triangles whose vertexes are configurations at three large times.  
The fact that the limit exists is a reasonable working assumption.
  (Note that we defined $f$ using the
correlation between the longest and the intermediate as the first
argument.) 

The function $f$ is time-reparametrisation invariant, 
associative $f(x,(y,z))=f((x,y),z))$, it has an identity and a zero, 
and it is bounded. Exploiting these
properties we showed that the most general $f$ is composed of pieces
satisfying either of the two forms: 
\begin{eqnarray}
f(x,y) &=& \jmath^{-1} \left(\jmath(x) \jmath(y) \right)
\; , 
\;\;\;\;\; \mbox{isomorphic to the product.} 
\label{isoproduct}
\\
f(x,y) &=& \min(x,y)
\; ,  
\;\;\;\;\; \;\;\;\;\; \;\;
\mbox{ultrametric,}
\label{ultrametricity}
\end{eqnarray}
This allows to classify every possible {\it Ansatz}.
Note that for $\jmath$ equal to the identity the first type of
function becomes simply $f(x,y)=x y$, hence the name.  It is also
possible to prove that the first kind of function (\ref{isoproduct})
is only compatible with the time scaling
\begin{equation}
C(t_2,t_1) = \jmath^{-1} \left( \frac{h(t_2)}{h(t_1)}\right)
\label{onescale}
\end{equation}
with $h(t)$ a monotonically growing function.  The 
dynamics of a given model can occur in two or more correlation
scales. 
In particular, for the SK model 
\begin{itemize}
\item[{\it i.}] 
$f$ is isomorphic to the product for correlation values in the 
stationary regime, $C>q_{ea}$, 
and $h(t)=e^{t/\tau}$. 
\item[{\it ii.}] 
$f$ is ultrametric for correlation values in the aging regime, $C<q_{ea}$.
\end{itemize}

Even though dynamic ultrametricity seems mysterious at first sight
there is a simple graphical construction that allows one to test
it. Take two times $t_3 > t_1$ such that $C(t_3,t_1)$ equals some
prescribed value, say $C(t_3,t_1)=0.3=C_{31}$.  Plot now $C(t_3,t_2)$
against $C(t_2,t_1)$ using $t_2$, $t_1 \leq t_2 \leq t_3$, as a
parameter. Depending on the value of $C_{31}$ with respect to $q_{\sc
  ea}$ we find two possible plots. If $C(t_3,t_1)> q_{\sc ea}$, for
long enough $t_1$, the function $f$ becomes isomorphic to the
product. Plotting then $C(t_3,t_2)$ for longer and longer $t_1$, the
construction approaches a limit in which $C(t_3,t_2) = \jmath^{-1}
(\jmath(C_{31})/\jmath(C(t_2,t_1)))$.  If, instead, $C_{31}< q_{\sc
  ea}$, in the long $t_1$ limit the construction approaches a
different curve. 

A preasymptotic scaling that leads to ultrametricity in the limit of
diverging times has been found by Bertin and Bouchaud in their study
of the dynamics of the critical trap model~\cite{Bebo}. Indeed, it is
simple to check that for any three correlations scaling as 
\begin{equation}
C(t,t') \sim \frac{\ln(t-t')}{\ln t'}
\label{eq:pre-ultra}
\end{equation}
relation (\ref{ultrametricity}) is recovered asymptotically. 

Ultrametricity in time is also very clear for a driven system satisfying
 (\ref{ludovic}). In the limit of small $\epsilon$ one can check that
$C(t_3-t_1)= \min[C(t_3-t_2),C(t_2-t_1)]$
 There is some evidence for it in the $4d$ Edwards-Anderson
(EA) model. In
$3d$ instead the numerical data does not support this
scaling~\cite{Picco}.  Whether this is due to the
short times involved or if the  asymptotic scaling is different in $3d$
is still an open question that will probably never be answered numerically or 
experimentally, as it was argued in Ref. \cite{Babeku} that 
time ultrametricity  would
take astronomic times to show up even if present asymptotically.

\subsection{Fluctuation-dissipation theorem (FDT)}

The analytic solution is such that, in the asymptotic limit in which
the waiting-time $t_w$ diverges after $N\to\infty$, the integrated
linear response approaches the limit
\begin{eqnarray}
\lim_{\stackrel{t_w\to\infty}{C(t,t_w)=C}}
\chi(t,t_w) = \chi(C)
\; 
\label{FDT}
\end{eqnarray}
when $t_w$ and $t$ diverge while keeping the correlation between them
fixed to $C$~\cite{Cuku94}.  Deriving this relation with respect to
the waiting time $t_w$, one finds that the opposite of the inverse of
the slope of the curve $\chi(C)$ is a parameter that replaces
temperature in the differential form of the FDT.  Thus, using
Eq.~(\ref{FDT}) one defines
\begin{equation}
T_{\sc eff}(C) \equiv -(\chi'(C))^{-1}
\label{teff_def}
\end{equation}
($k_B=1$), 
that can be a function of the correlation. Under certain circumstances
one can show that this quantity has the properties of a
temperature~\cite{Cukupe}.

One of the advantages of this formulation is that, just as in the
construction of triangle relations, times have been ``divided away''
and the relation (\ref{FDT}) is invariant under the reparametrisations
of time (\ref{eq:RpG}). Moreover, the functional form
taken by $\chi(C)$ allows one to classify glassy systems into sort of
`universality classes'.

Equation~(\ref{FDT}) is easy to understand graphically.  Let us take a
waiting time $t_w$, say equal to $10$ time units after the preparation
of the system (by this we mean that the temperature of the environment
has been set to $T$ at the initial time) and trace $\chi(t,t_w)$
against $C(t,t_w)$ using $t$ as a parameter ($t$ varies between $t_w$
and infinity). If we choose to work with a correlation that is
normalised to one at equal times, the parametric curve starts at the
point $(C(t_w,t_w)=1,\chi(t_w,t_w)=0$) and ends in the point
$(C(t\to\infty,t_w)\to 0,\chi(t\to\infty,t_w)= \overline\chi$).  Now,
let us choose a longer waiting time, say $t_w=100$ time units, and
reproduce this construction.  Equation~(\ref{FDT}) states that if one
repeats this construction for a sufficiently long waiting time, the
parametric curve approaches a limit $\chi(C)$.

In the SK model one finds
\begin{eqnarray}
T \; \chi(C) = 
\left\{
\begin{array}{ll}
1-C \; , &\qquad C>q_{ea}
\\
1-q_{ea} + (q_{ea}^2-C^2) \; , &\qquad C<q_{ea}
\end{array}
\right.
\end{eqnarray}
This result corresponds to having a succession of temporal scales
each one with an effective temperature, $T_{\sc eff}(C)$.

The question as to whether this behaviour 
strictly applies to the finite dimensional case remains open.
Fluctuation-dissipation violations, {\it i.e.} the existence of $T_{eff} \neq T$
in the infinite waiting-time limit
do exist, for instance
in systems with growing domains, but we still have no examples
in which we are certain that $T_{eff}$ stays 
bounded away both from $T$ and from infinity t, in the large waiting-time
limit
(the latter being the case for coarsening models).  
Numerical simulations in finite dimensional spin-glass and structural 
glass models show a clear deviation from FDT during the out of equilibrium 
relaxation but these are obtained for finite times and although 
these times are relatively long it is hard to extract the truly asymptotic 
limit.

\subsection{Functional dynamic order parameter }

In~\cite{Cuku93} a set of generalised susceptibilities
\begin{eqnarray}
I^r(t)
&\equiv&
\lim_{N\rightarrow\infty}
\frac{r!}{N^r}
\sum_{i_1 < \dots < i_r}
\left[
\frac{\delta
\langle \,
s_{i_1}(t) \dots s_{i_r}(t)
\, \rangle
}
{
\delta
h_{i_1 \dots i_r}
}
\right]_{h=0}
\nonumber\\
&=&
r
\int_0^t dt' \, C^{r-1}(t,t') \, R(t,t')
\; ,
\label{int}
\end{eqnarray}
and their generating function
$P_d(q)$
\begin{equation}
1-
\lim_{t \rightarrow \infty} \lim_{N \rightarrow \infty}
\frac{r!}{N^r}
\sum_{i_1 < \dots < i_r}
\left[ 
\frac{\delta
\langle \, 
s_{i_1}(t) \dots s_{i_r}(t)
\, \rangle
}
{
\delta
h_{i_1 \dots i_r}
}
\right]_{h=0}  
=
\int_0^1 dq' \, P_d (q') \, {q'}^r
\; .
\label{aa}
\end{equation}
were introduced. The physical meaning of $P_d(q)$ is clear.
If the order of large system size and long time limits are 
reversed the generating 
functional becomes Parisi static functional order parameter.

The analytic solution of the non-equilibrium dynamics of the SK model
is such that $P_d(q)=P(q)$ -- even if the physical situations that
these two order parameters describe are very different~\cite{Cuku94}.
All generalised susceptibilities converge, then, to the equilibrium
values~\cite{Frme}.
 This result suggests that the landscape the SK dynamics visits
at different long times is similar to the one that characterises the
equilibrium pure states (though with finite barriers separating the
traps visited dynamically). Although at long but finite times with
respect to $N$ the SK model explores regions of phase space that it
will eventually leave never to return, some geometrical properties of
these regions coincide with those of the equilibrium states.

A similar conclusion was reached in a slightly different context in 
Ref.  \cite{Bacukupa95}.  The quantities studies there were
the staggered auto-correlation and 
linear response:
\begin{eqnarray}
C(\lambda;t_1,t_2)
&\equiv&
\langle
\sigma_\lambda(t_1) \; \sigma_\lambda(t_2)
\rangle
=
\sum_{ij}
\langle \lambda | i \rangle \;
\langle \lambda | j \rangle \;
\langle \sigma_i(t_1) \; \sigma_j(t_2) \rangle
\;.
\label{stagcorr}
\\
R(\lambda;t_1,t_2)
&\equiv&
\left\langle 
\frac{\delta\sigma_\lambda(t_1)}{\delta h_\lambda(t_2)} \right\rangle
=
\sum_{ij}
\langle \lambda | i \rangle \;
\langle \lambda | j \rangle \;
\left\langle \frac{\delta\sigma_i(t_1)}{\delta h_j(t_2)} \right\rangle
\; ,
\label{stagresp}
\end{eqnarray}
where $\lambda$ denotes the eigenvalues of the $N \times N$ random
matrix $J_{ij}$ associated with the eigenvectors $| \lambda \rangle$.
$|\sigma(t) \rangle$ is the time-dependent $N$-dimensional vector of
spins, $\sigma_i(t) \equiv \langle i | \sigma(t) \rangle$, and
$\sigma_\lambda(t) \equiv \langle \lambda | \sigma(t) \rangle$ are the
staggered spin states. We showed that the staggered auto-correlation
distribution $C(\lambda,{\cal C})$, between two large and widely
separated times $t_1$ and $t_2$ chosen such that $C(t_1,t_2)={\cal C}\leq
q_{EA}$ coincides with the static one computed with configurations
belonging to two equilibrium states with mutual overlap ${\cal
  C}$. Moreover, if one stores the configuration at times $t_2$ and let
the system evolve up to a time $t_3$ such that again $C(t_2,t_3)={\cal
  C}$ one obtains the same form for the staggered correlation
$C(\lambda,t_2,t_3)$.  (Note, however, that due to the system's slowing
down, $t_2-t_1<t_3-t_2$ if ${\cal C}<q_{EA}$.)  

The one-time quantities (e.g. nonlinear susceptibilities and
 staggered magnetisation) derived from the dynamics of the SK model 
thus coincide with those calculated in equilibrium. This fact, though rather
surprising for a mean-field model, has been derived under certain assumptions
for finite-dimensional models \cite{Franz}.

\begin{figure}[h]
\begin{center}
\epsfig{file=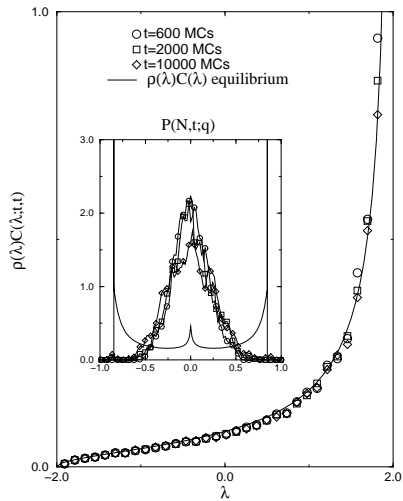,width=6cm}
\end{center}
\caption{Staggered correlation at different
times given in the key.  Inset: the overlap
distribution at the same times. The full lines correspond
to the analytic prediction~\cite{Bacukupa95}.}
\label{fig:Stag}
\end{figure}

At the mean-field level,  this coincidence  holds for
models that do not have a `threshold' level below which the system
cannot penetrate in finite times with respect to $N$. Examples are
{\it e.g.} the SK model and the model of a manifold in a random
potential with long-range correlations~\cite{Frme,Cule}, both having a
continuous set of correlation scales and being solved 
by a full replica symmetry breaking (RSB) scheme at the static 
level. Instead, there is no reason why the free-energy landscape 
explored dynamically should resemble the static one in models with a 
threshold, such as the $p$-spin spherical model~\cite{Cuku93}
(that is characterised, statically by a one-step level 
of RSB).

Let us remark that the good agreement between the numerical
calculation of $C(\lambda,t,t)$ for large $t$ and the static
distribution $C(\lambda)$, see Fig.~\ref{fig:Stag}, constitutes a
rather detailed test of the solution of the out of equilibrium
dynamics for this model.
 Further studies of the organisation 
of metastable state and their relevance to the out of equilibrium 
relaxation of the SK model appeared in~\cite{metastable}.

\subsection{ Temperature cycling protocols}

 A means to study the dynamics in the glassy phase in more
detail
 consists in following the evolution of the sample under a
complicated
 temperature history.  The protocols that have been more
commonly used
 include temperature and field cycling within the low
temperature
 phase~\cite{Delta-T-exp}.
 Different types of glasses
show rather different responses to the
 change in external
parameters.  {\em Spin}-glasses show the puzzling
 phenomenon of
reinitialisation of aging following a decrease in
 temperature,
combined with the recall of the situation attained before
 the
downward jump when the original high temperature is
restored. Remarkably, when similar protocols were applied to {\em
structural} glasses, e.g. in dielectric constant measurements of
glycerol by Leheny and Nagel, no substantial reinitialisation was
observed~\cite{Leheny}. Experiments in {\it dipole} glasses display
an
 ``intermediate'' behaviour in the sense that a temperature
cycling provokes strong asymmetric results, as in spin-glasses,
while they also
 present very strong dependencies on the cooling rate,
a property that
 is not observed in spin-glasses though is very
common in structural
 glasses~\cite{Alberici}.
 
 Mean-field
inspired researchers have interpreted the experimental
 results of
temperature cycling experiments using a hierarchical dynamic
 picture
inspired by the organisation of
 equilibrium states in the Parisi
solution of the SK model. In this
 picture one assumes that
spin-glasses have a large number of
 metastable states that are
organised in a hierarchical fashion just
 like the equilibrium
states.  It is then proposed that the system is
 composed of
(independent) subsystems whose dynamics is given by the
 wandering in
such a landscape~\cite{aging-mean-field}.  An average over subsystems
has to be
 invoked in order to
 obtain smooth results as observed in
experiments.
 Instead, droplet picture~\cite{Fisher-Huse} oriented
researchers found the outcome of the same 
 experiments unequivocal
evidence for their favourite theory~\cite{aging-droplets}.

The outcome of temperature cycling experiments in spin-glasses can be
understood within the dynamic solution of the SK model. Moreover,
the reasons why these effects should be hardly observable in
spin-glasses at very short times such as are inevitably involved in
simulations, and they are absent in structural glasses are clear
within the analytic solution to the
 dynamics of mean-field models~\cite{Cuku99}.

The key dynamic property to explain the outcome of these experiments
is the sharp separation of correlation-scales in the asymptotic
waiting-time limit. Take a fixed (but very long) waiting time and let
the system evolve further. Imagine at the subsequent time $t$ the
self-correlation is $C=q_0\leq q_{ea}$.  The decay below this value
needs a time-delay that is infinitely longer than $t-t_w$. In other
words, at any times $t$ and $t_w$ the correlation (and linear
response) can be separated in two terms, a fast and a slow one, such
that for all time delays such that $C_{fast}$ changes $C_{slow}$ is
fixed while, instead, if $C_{slow}$ varies $C_{fast}$ has reached its
limiting value.

The effect of a temperature jump is then very different on the fast
and the slow scales. The easiest way to visualise it is to use
Fig.~\ref{sketch}~(a).  Upon changing the external temperature the FD
plot is modified by changing the slope of the linear part representing
the equilibrium FDT result and, in consequence, the intercept of the
line with the curve part that remains unchanged under the Parisi-Toulouse 
(PaT)
hypothesis~\cite{Pato} that consists in two assertions: 
\begin{itemize}
\item[{\it i.}]
$\chi(C)$ is independent of $T$ and $H$ in the aging regime;
\item[{\it ii.}]
$q_{ea}$ only depends on $T$ and $q_0$ only depends on $H$.
\end{itemize}
The near
temperature-independence of $\chi(C)$ in the aging regime of the SK
model has been argued at the level of Parisi static solution and
carries through to the non-equilibrium relaxation due to
$P_d(q)=P(q)$.  It is also a very good approximation in the $4d$ EA
model as checked numerically. As far as we know, there are no
tests of this hypothesis in the 3D case.

For temperature
$T$ the thin solid line in Fig.~\ref{sketch}~(a) represents the equilibrium
result. In the figure we show the FDT part for a different
temperature that we called $T_g(H)$ for the purposes of the discussion
of experimental measurements of FDT violations. Here we interpret
$T_g(H)$ just as a higher temperature and $q_0(H)$ as its
Edwards-Anderson parameter (there is no applied field). For
temperature $T_g(H)$ the equilibrium result is the dashed straight
line.  The thick black curved line is the same at both temperatures.
The effect of a temperature change is then quite different on the slow
and fast correlation scales. It corresponds to the clockwise or
anticlockwise motion of the straight line part of the plot.
  The slow scales are just modified by
a time-parametrisation (\ref{eq:RpG}), 
independently of the jump being positive or
negative.  The scales between $q_0$ and $q_{ea}$ are instead created
or destroyed (restarted or erased) by the negative and positive
temperature jumps. This intuitive idea -- very close to the one 
put forward in the hierarchical explanation of temperature jump 
experiments -- can be made precise and implemented in analytic 
calculations~\cite{Cuku99}.

An argument along the same lines allows one to explain the outcome
of field jump experiments. 

The phenomenology of structural glasses is described by models of the
$p$-spin type that realise the random first order transition
scenario~\cite{KTW}. The aging dynamics of these systems occurs in only one
time-scale, typically described by a simple $t/t_w$ scaling. In these
cases the argument described above does not apply (the decay of the
correlation below any value $C<q_{ea}$ is not infinitely slower than
the one that occurred before). This yields a theoretical justification
of the fact that the outcome of temperature variation experiments in
other glassy systems are quite different from the ones in
spin-glasses.

\subsection{Fluctuations: towards a Sigma Model approach}

Observables in finite size systems fluctuate. A theory for the
disorder-averaged, noise-induced dynamic fluctuations of finite
dimensional glassy systems was proposed in \cite{Ch-etal}.  These
fluctuations are not induced by the particular realisation of quenched
disorder but should be generated dynamically in models in which
time-parametrisation invariance develops asymptotically.

The natural counterpart to the coarse-grained local correlations
and responses in finite dimensional models is, for a fully 
connected model, the global quantity itself. The latter
fluctuates if the fully-connected system has a finite size.

One of the main consequences of the time-parametrisation
invariance theory of fluctuations is that the fluctuations
in the fluctuation-dissipation relation in the aging regime 
should distribute along the global $\chi(C)$ curve. 
In this Subsection we review the analysis of such 
fluctuations obtained numerically for finite size SK models. 

\vspace{.2cm}

\noindent
{\em Finite size fluctuations of global quantities.}

\vspace{.2cm}

If one wishes to show that a given system with a broken symmetry
tends to behave like
a Sigma model in some limit, what one has to do is to plot the fluctuations
of the `radial' variables that are left invariant by the group, and
 check that they become vanishingly small compared to those of the `angular'
variables generated by the group itself.
In precisely that spirit, 
in order to show that the system's fluctuations explore
preferentially the (almost) flat directions generated by 
reparametrisation invariance, one can plot the fluctuations
of correlations and responses in such a way as to show that 
fluctuations of quantities
left invariant by this group (the departures from a $\chi$ vs $C$ curve)
become negligible with respect to fluctuations generated
 by reparametrisations (along the $\chi$ vs $C$ curve).

\begin{figure}[h]
\centerline{
\psfig{file=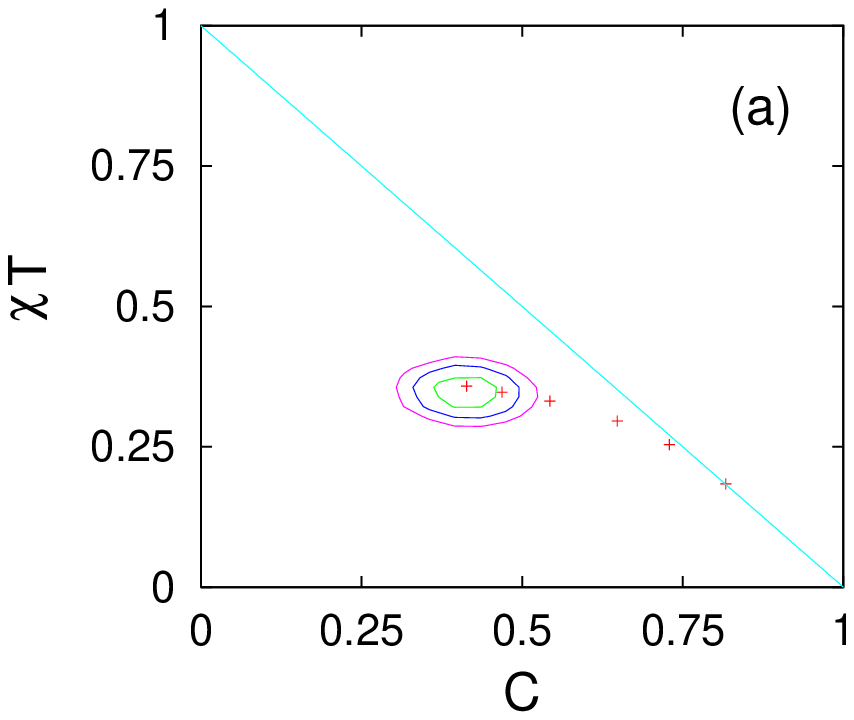,width=6cm}
\psfig{file=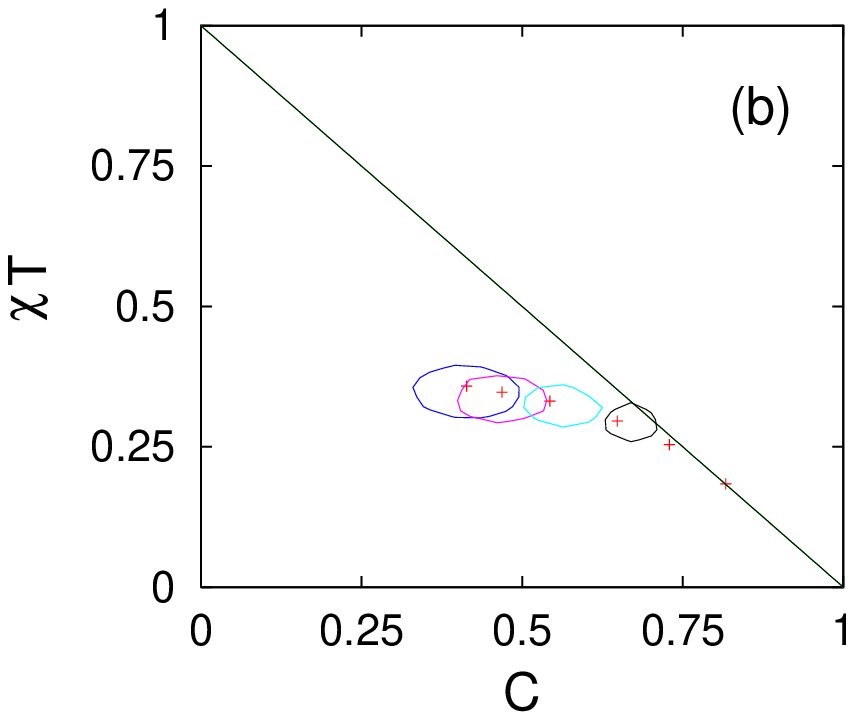,width=6cm}
}
\vspace{0.25cm}
\caption{Projection of the joint probability distribution 
function (PDF) for the global
susceptibility and correlations of the SK model with $N=512$ and
$\beta=2.5$. The strength of the applied field is $\eta=0.25$. A
coarse-graining over time is done using $\tau=2$ for $t_w=64$ {\sc
mc}s and $t=65,70$ {\sc mc}s, and $\tau_t=2$, $4$, $8$, and $16$ {\sc mc}s
for $t=128$, $256$, $512$,and $1024$ {\sc mc}s.  The crosses indicate values
averaged over the distribution, the straight line is the prediction
from the FDT. In panel (a) the contour levels are chosen at
heights corresponding to $95\%$, $90\%$, and 82\% of the maximum
in the PDF for the global correlations evaluated at $t_w=64$
{\sc mc}s and $t=1024$ {\sc mc}s. In panel
(b) the contour levels  are at $90\%$ of the maximum and they correspond
to the PDFs calculated at $t_w=64$ {\sc mc}s and $t=128,256$,
$512,1024$ {\sc mc}s from right to left. Figure taken from~\cite{Cachcuigke03}.}
\label{fig:sk-finite-size}
\end{figure}

 Figure~\ref{fig:sk-finite-size} shows 
the level curves of the joint probability
of the  global susceptibility and  correlation of an SK model for
a  system with
$N=512$ at $T=0.4$.
The distribution functions were obtained using $10^5$ pairs
$(C(t,t_w),\chi(t,t_w))$
with $10^4$ different noise
histories, and repeating this procedure
with $10$ different realisations of disorder.
The straight line represents the equilibrium FDT.
We see that the probability distribution
is peaked on  the global 
$T \chi(C)$ curve.  Thus, we conclude that different histories 
tend to be affected by random  time reparametrisations, just as a Sigma model
tends to fluctuate along the angles spanned by the group.
Very similar results have been obtained in the $3d$ EA model~\cite{Cachcuigke03}.

\vspace{.2cm}

\noindent
{\em Fluctuations in the noise-averaged local quantities}

\vspace{.2cm}

\begin{figure}[h]
\centerline{
\psfig{file=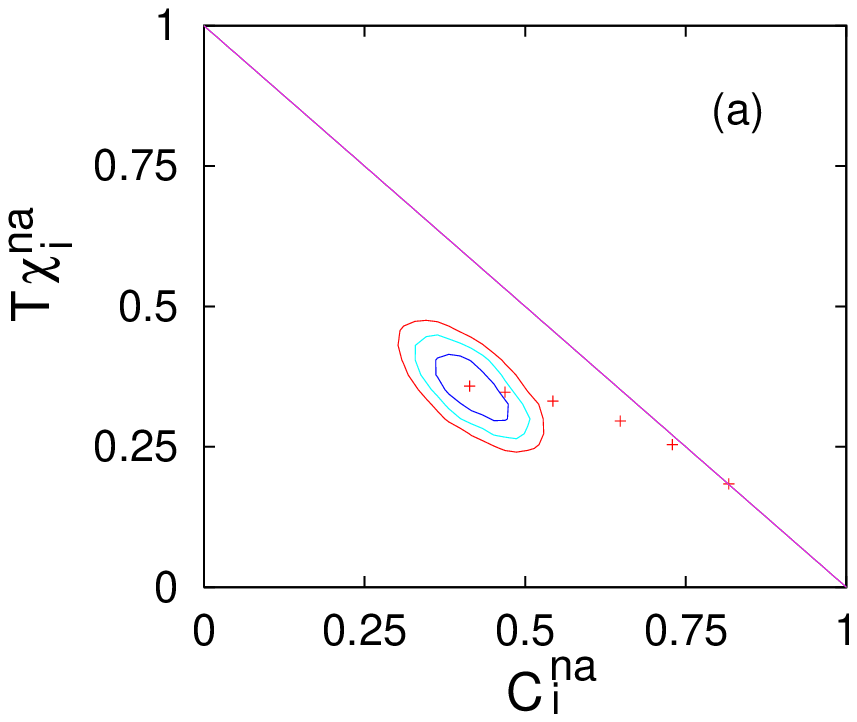,width=6cm}
\psfig{file=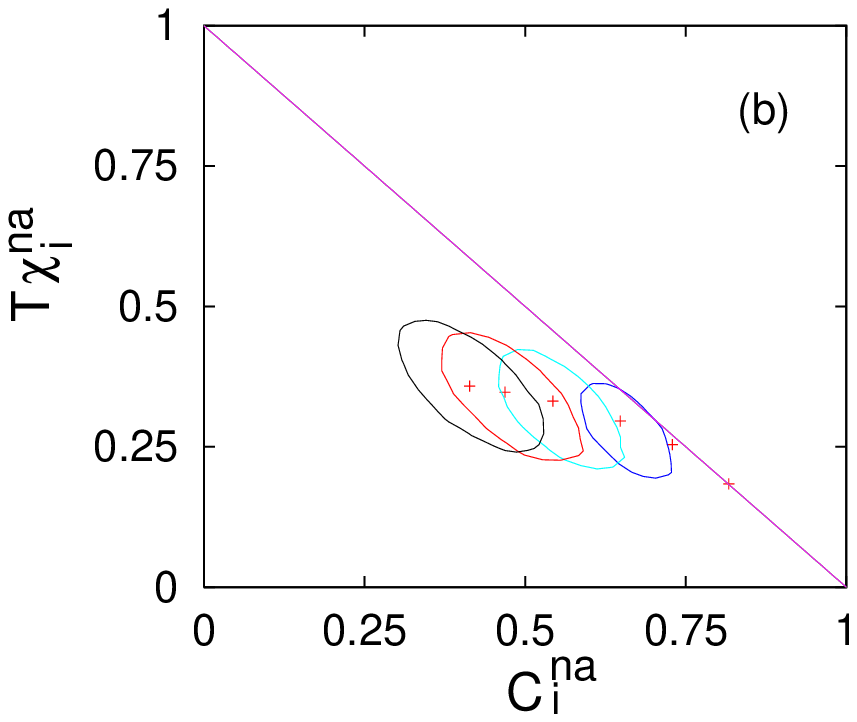,width=6cm}
}
\vspace{0.25cm}
\caption{Projection of the 
joint PDF for the noise-averaged
``local'' susceptibilities and correlations
of the SK model with $N=512$ and $\beta=2.5$. The strength of the 
applied field is $\eta=0.125$. The coarse-graining times $\tau$ are 
chosen as in Fig.~\ref{fig:sk-finite-size}. The crosses indicate 
values averaged over the distribution, the straight line
is the prediction from the FDT. In panel (a) the contour levels are chosen
at heights corresponding to $90\%, 85\%, 80\%$ and they correspond
to times $t_w=64$ {\sc mc}s and $t=1024$ {\sc mc}s. In panel (b) the contour
levels  are at $80\%$ and they correspond to the joint 
PDF at $t_w=64$ {\sc mc}s and $t=1024$ {\sc mc}s.
Figure taken from~\cite{Cachcuigke03}.}
\label{fig:sk-noise-averaged}
\end{figure}

The existence of  soft modes for time-parametrisation is a feature 
of slow dynamics, quite independent of the presence of quenched disorder.
In order to stress this point,  Fig.~\ref{fig:sk-noise-averaged}
shows the distribution over sample realisations  of correlations and response
function Fig.~\ref{fig:sk-finite-size}, the data for each sample being averaged
over the noise.  The orientation of the contour levels
does not follow the $T\chi(C)$ curve but, instead, it is approximately
parallel to the FDT straight line.
Clearly, the sample-to-sample variations have nothing to do with
the reparametrisation invariance, which is a dynamic effect.
Again, very similar results are obtained in the $3d$ 
EA model. 

\section{Experiments}
\label{sec:experiments}

The SK model is, undoubtedly, the mean-field model of spin-glasses.
One would then like to confront its dynamic behaviour to the one 
observed experimentally. In this Section briefly do this
by discussing some salient
experimental results. 
  
\subsection{Isothermal aging}

Several groups studied the relaxation of spin-glasses using 
ac-susceptibility and dc magnetisation measurements. The experimental 
data -- as well as the numerical data from simulations of the 
$3d$ EA model -- are rather well described by a much simpler time-dependence
than the one found in the SK model. More precisely, neither the ultrametric 
 relation (\ref{ultrametricity}) 
nor its pre-asymptotic form (\ref{eq:pre-ultra})
fit the data satisfactorily. Instead, the data are rather accurately described 
by a two-scale scenario with an aging regime characterised by an 
enhanced power law, $h(t) \propto e^{\ln^a(t/t_0))}$, with $a\sim 2$,  
that weakly deviates from a simple power~\cite{Vincent}. Note that this 
scaling is also different from the one predicted by the droplet 
model~\cite{Fihu}.

\subsection{ Temperature cycling experiments}

Temperature cycling experiments are amongst the most striking and 
beautiful ones made with spin-glasses.
As mentioned above, the SK model responds to temperature cycling
in a manner qualitatively very close to that of experiments.
What remains a mystery however, is that  the memory effects 
in the SK model arise thanks to the existence of
many widely-separated timescales, while
in the experimental system there seems to be only a single scale.

\subsection{Fluctuation-dissipation relation}

\vspace{.2cm}

{\em Direct measurements}

\vspace{.2cm}

Deviations from the FDT should be tested by measuring the dynamic
induced and spontaneous fluctuations of a chosen observable using the
same experimental device. In a remarkable series of experiments,
H\'erisson and Ocio carried out such a study
focusing on the magnetisation of an isolating spin-glass
sample~\cite{Heoc}.  Their results for correlation and linear response
functions are compatible with the two-scale scenario and the enhanced
power law aging time-scale. The direct comparison between integrated
linear response and correlation function yields an FD plot that has
been interpreted by the authors as being similar to the curved shape
of the SK model. It should be stressed, however, that this
interpretation is inconsistent with the understanding of the
fluctuation-dissipation deviations as being related to the existence
of effective temperatures~\cite{Cukupe}, since the whole decay 
occurs
in a single timescale. Moreover, in our opinion, the
resulting FD plot cannot be really distinguished from a broken
straight line which is consistent with a two-scale scenario and the
effective temperature interpretation. The same {\it proviso} applies to the 
numerical data for the $3d$ EA model. 

\vspace{.2cm}

\noindent
{\em Zero-field and field-cooled magnetisation}

\vspace{.2cm}

An indirect study of the fluctuation-dissipation relation using
several sets of experimental data obtained from various samples was
presented in~\cite{Cugrkuvi99}.  The proposal, motivated by
discussions with DS Sherrington during a visit to the University of
Oxford, is to use the well-known difference between the field-cooled
and zero-field cooled magnetisation in the low $T$ phase to infer the
deviation from the fluctuation-dissipation relation between
spontaneous and induced fluctuations.

The approach uses a dynamic extension of the Parisi-Toulouse
(PaT) approximation~\cite{Pato} that we explained above. 
The PaT approximation allows
us to estimate the $C$-dependence of the susceptibility using {\it
  exclusively} response results, thus circumventing the difficulties
inherent to noise measurements. Deviations from the Curie-Weiss 
law due to a non-vanishing average of the exchange coupling in 
real spin-glasses were also taken into account. 

\begin{figure}[h]
\vspace{2cm}
\begin{center}
\epsfig{file=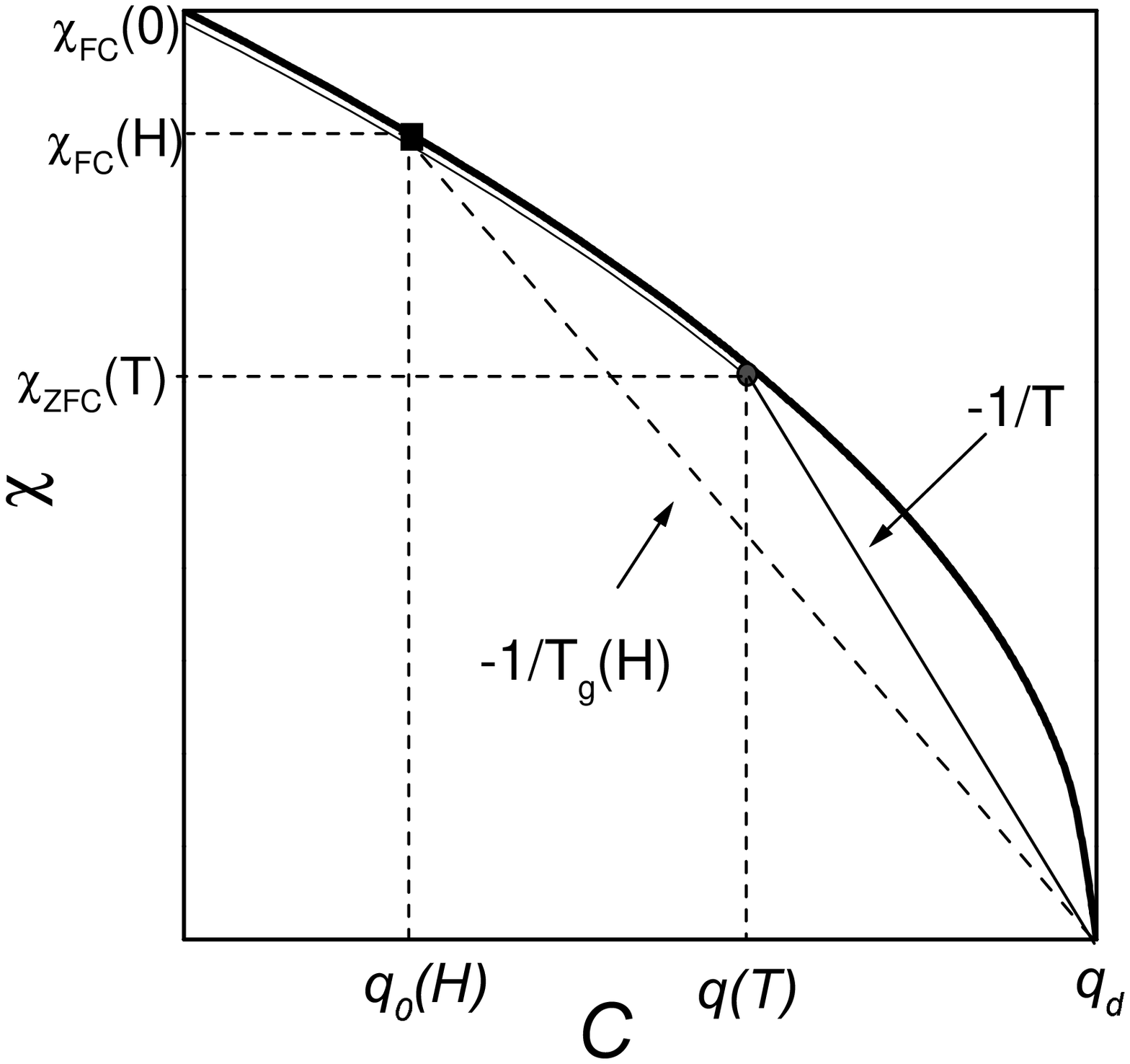,width=6cm}
\epsfig{file=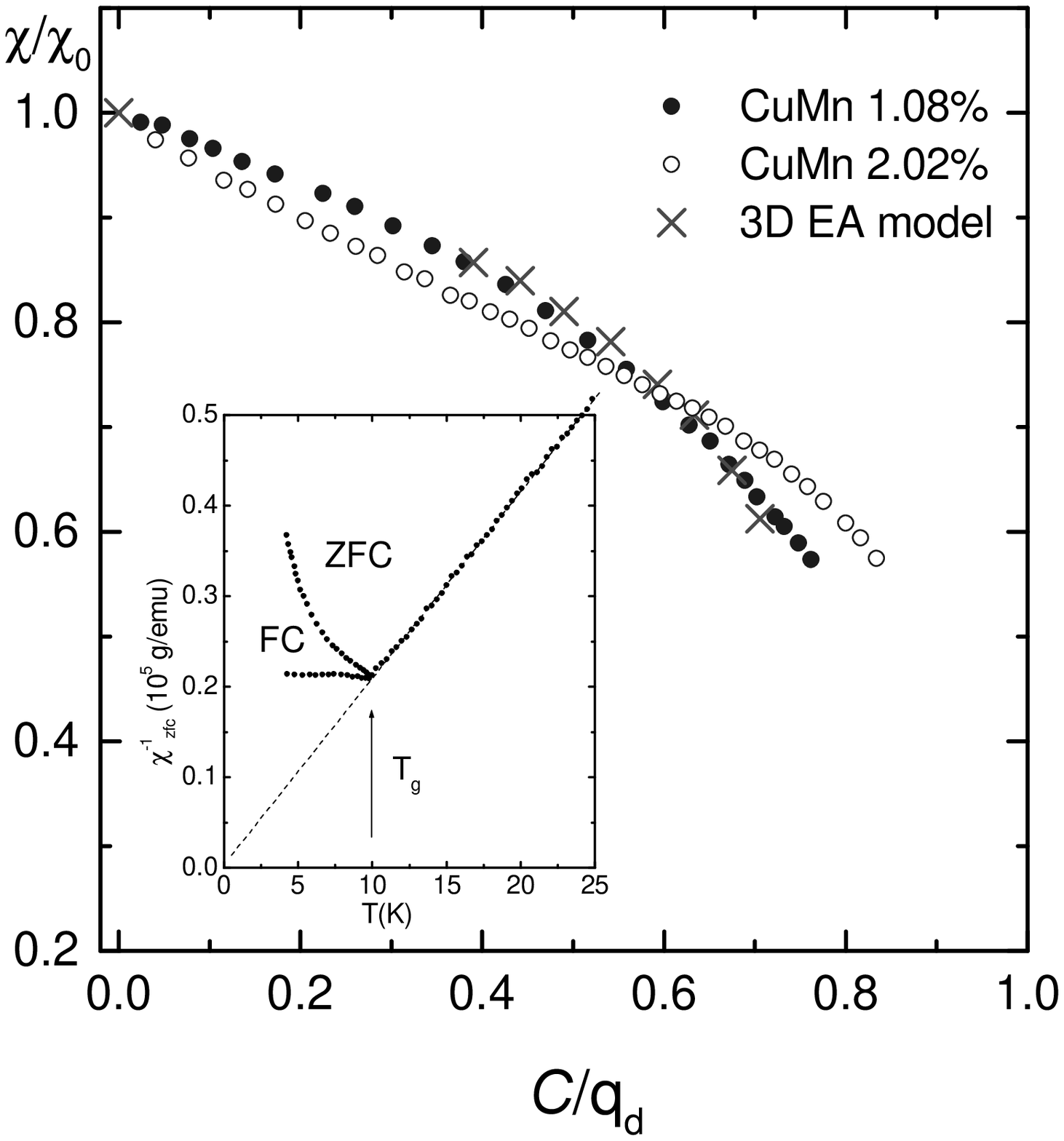,width=6cm}
\end{center}
\caption{Left: sketch of the $\chi$ vs $C$ plot. The thick curve
  represents the master curve ${\chi}_{ag}(C)$ that, within the PaT
  approximation, is temperature and field independent.  The thin
  straight line has slope $-1/T$ ($T<T_g$) and represents
  Eq.~(\ref{chizfc}). The dashed straight line has a slope $-1/T_g(H)$
  and joins $(q_d,0)$ to $(\chi_{\sc fc}(H),q_0(H))$.  Right:
  $\chi_{ag}(C)$ plot for CuMn at 1\% and 2\%. 
The vertical axis is normalised by the
  susceptibility at the critical temperature in zero field
  ($\chi_0$). The horizontal axis is normalised by $q_d$.  The crosses
  are numerical results for the $3d$ EA model \cite{Juan3DEA}. The inset
  shows the inverse {\sc fc} and {\sc zfc} susceptibilities as
  functions of temperature.  Figure taken from Ref.~\cite{Cugrkuvi99}.}
\label{sketch}
\end{figure}

The strategy is to use data taken under $T$ and $H$ conditions such
that the system is at the limit of validity of FDT, {\it i.e.}
$C(t,t_w) \sim q_{ea}$. The point $\{ q_{ea},\chi(q_{ea})\}$ is the 
intersection
between the straight part (FDT regime) and the aging part of $\chi(C)$ where
\begin{eqnarray}
\chi(q_{ea}) = \lim_{t-t_w\to\infty} \lim_{t_w\to\infty}
\chi(t,t_w) = \frac{1}{T} (q_d - q_{ea})
\label{chizfc}
\end{eqnarray}
and to associate $\chi(q_{ea})$ to the zero-field cooled
susceptibility $\chi_{zfc}$ measured experimentally.  $q_d$ is the
equal time correlation that is not necessarily one but can be obtained
from $\lim_{T\to 0}q_{ea}$.  The locus of the points obtained by
varying $T$ spans a {\it master curve} $\chi_{ag}(C)$ which, by the
PaT hypothesis, is field and temperature independent. At a given
working temperature $T$ the actual $\chi(C)$ curve consists of a
straight line with slope $-1/T$ joining $(q_d,0)$ and
$(q_{ea},\chi(q_{ea}))$ and a second part given by $\chi_{ag}(C)$. The
method of construction is explained in \cite{Cugrkuvi99} and it is
illustrated in Fig~\ref{sketch}.  A complementary argument uses 
the field-cooled magnetisation to construct $\chi_{ag}(C)$ by spanning
$\chi_{FC}(H)=1/T_g(H)(q_d-q_0(H))$ as a function of $H$.

The analysis is most reliable for CuMn, a system in which the
Curie-Weiss law as well as the PaT approximation are very well
verified.  Figure~\ref{sketch} shows the $\chi_{ag}(C)$ curve
determined using the {\sc zfc} data of Nagata {\it et al} for two
concentrations~\cite{Nagata}. There are no experimental points
for $C/q_d>0.8$ that correspond to rather low temperatures.  We know
however that $\chi(C)$ tends to zero as $C \rightarrow q_d$
since $\chi_{\sc zfc}(T=0)=0$.  In addition, the slope
$d\chi/dC$ should be infinite at $C=q_d$ so that $q=q_d$ only
at $T=0$.  The validity of the hypotheses can be judged by the inset
where we show the temperature dependence of the inverse susceptibility
for the 1.08\% compound.  A Curie-Weiss law with $\theta \approx 0$
holds accurately for all $T \ge T_g$. The $T$-independence of
$\chi_{\sc fc}$ required by the PaT approximation is also well
verified below the transition. The same is true for the 2.02\% sample.
For comparison, we also show the curve
$\chi(C)$ for the $3d$ EA model, at $T=0.7 (<T_g)$ and $H=0$,
obtained numerically in Ref.~\cite{Juan3DEA}.  The agreement between
the numerical results and the experimental data for the 1.08\% sample
is remarkable. It may be fortuitous, however, since the results for
the 2.02\% sample deviate from it. In fact, one must note that
$\chi(C)$ is not a universal function. For example, it depends
on the details of the Hamiltonian (Heisenberg, Ising and, in general,
the level of anisotropy) even at the mean-field level.  Thus, there is
no reason to expect universality in real systems.

Note that in real samples a spin-glass transition in a field may not
exist.  However, even if this were the case, the system should remain
below a slowly time-dependent {\it pseudo} de Almeida-Thouless (AT)
line for still relatively long times: it ages and behaves as a true
(out of equilibrium) glass with a non-trivial $\chi(C)$ that would
eventually become a straight line with slope $-1/T$.  

Another important issue is the asymptotic ($t_w\to\infty$) 
form of the $\chi_{\sc ag}(C)$ curve. 
Even if the system never equilibrates,
the $\chi_{\sc ag}(C)$ curve may still be a very slowly
varying function of $t_w$, eventually
reaching a form different from that  observed 
experimentally. One is not in a position to discard this possibility.

As was mentioned above, equilibrium and large-time non-equilibrium 
one-time quantities are expected (under certain assumptions) to 
coincide~\cite{Franz}.
If these hypotheses are warranted for spin glasses, 
the slope of the dynamic $\chi(C)$, for an 
infinite system in the large-$t_w$ limit should coincide
with the static $x(q)$ as defined by the probability of 
overlaps of configurations
taken with the Gibbs measure -- the connection being established
through the generalised susceptibilities as explained above.  
The determination of a non-trivial  $\chi(C)$ would thus imply
a non-trivial $x(q)$ and would hence validate Parisi's solution.
The problem, as is usual in these systems, lies in the fact that the dynamics,
even in the  experimental case, are confined to quite short times.

\section{Conclusion}

 From the dynamic point of view, the Sherrington-Kirkpatrick model
has some aspects in which it resembles experimental systems, and some
 in which it does not. 
Amongst the resemblances, as we have seen, one can count the 
fluctuation-dissipation characteristics and the remarkable
temperature-cycling properties of memory loss and recovery.

 On the other hand, there is the inescapable 
fact that one does not see  any
evidence of dynamic ultrametricty either in experimental systems or
in their numeric  counterparts, while these are easily observable
in simulations of the SK model~\cite{Ba}.  Even assuming that at longer times
the separation of timescales would develop, this can be estimated to
happen not before astronomic times~\cite{Babeku}.

As mentioned in the previous section, a non-trivial  $\chi(C)$ at very long
times is an indication of a nontrivial Parisi function. The problem is that
experimentally accessible times are not that long --- coherence length scales
  of around twenty are estimated in the best of cases.
An apparently nontrivial   $\chi(C)$ that would eventually become trivial --
 or an apparent  Almeida-Thouless  line disappearing at long times -- 
 would be an example of  a phenomenon that is
a permanent source of confusion: finite size systems in equilibrium, and
 infinite-size systems at short times, tend to have a pre-asymptotic behaviour
that looks  qualitatively mean-field like.
This tendency can be judged as positive -- because the mean-field picture is
then a qualitative  model of what we see in practice, or negative
 -- because it does not allow us to distinguish properly between theories.


\vspace{0.2cm} 

The Sherrington-Kirkpatrick model was originally designed as a 
toy model of spin glass, that would serve as a straightforward, practical
 starting point.  Fortunately for us, this expectation  
 proved unfounded,
as over thirty years of surprises and amusement have shown.

\vspace{0.5cm} 

\noindent
{\bf Acknowledgements}
LFC is a member of Institut Universitaire de France.

\end{document}